# Optical patterning of trapped charge in nitrogen-doped diamond

Harishankar Jayakumar[1], Jacob Henshaw[1,2], Siddharth Dhomkar[1], Daniela Pagliero[1], Abdelghani Laraoui[1], Neil B. Manson[3], Remus Albu[1], Marcus W. Doherty[3], and Carlos A. Meriles[1,2,*]

The nitrogen-vacancy (NV) centre in diamond is emerging as a promising platform for solid-state quantum information processing and nanoscale metrology. Of interest in these applications is the manipulation of the NV charge, which can be attained by optical excitation. Here we use two-color optical microscopy to investigate the dynamics of NV photo-ionization, charge diffusion, and trapping in type-1b diamond. We combine fixed-point laser excitation and scanning fluorescence imaging to locally alter the concentration of negatively charged NVs, and to subsequently probe the corresponding redistribution of charge. We uncover the formation of spatial patterns of trapped charge, which we qualitatively reproduce via a model of the interplay between photo-excited carriers and atomic defects. Further, by using the NV as a probe, we map the relative fraction of positively charged nitrogen upon localized optical excitation. These observations may prove important to transporting quantum information between NVs or to developing three-dimensional, charge-based memories.

The future realization of quantum technologies in the solid state largely relies on our ability to identify physical platforms compatible with the local manipulation of quantum information and subsequent transport to remote processing nodes. One possibility — presently attracting widespread interest — is the use of select color centres and paramagnetic defects in semiconductors, including silicon[1,2], silicon carbide[3-5], diamond[6], and yttrium aluminum garnets[7]. Of interest herein is the negatively charged nitrogen vacancy centre (NV⁻), a paramagnetic defect formed by a substitutional nitrogen immediately adjacent to a vacant site in the diamond lattice[8]. Over the last decade, extensive studies have demonstrated the use of the NV⁻ (and neighboring nuclear and electronic spins) as a valuable resource to locally store and process quantum information[6]. On the other hand, chains of spin-active defects[9,10] and spin-photon entanglement[11] are being explored as possible routes to communicate distant NVs. In these architectures, quantum state transport is mediated by the dipolar couplings between the spins in the chain or by spin-encoded photons propagating through a photonic network, respectively.

An alternate approach to defect networking could rely on the use of electrons photo-generated from separate, non-interacting NVs, as proposed recently[12]. Indeed, prior work on the ionization dynamics of individual defects demonstrates that the NV charge state can be controlled optically[13,14], thus prompting one to view the nitrogen-vacancy centre alternatively as a source of photons or charge carriers. Regrettably, no prior work exists on the transport of charge carriers between NVs.

Further, the charge exchange between NVs and nitrogen donors — often invoked when explaining the origin of the NV⁻ excess electron — has not yet been experimentally probed.

Here we use two-color fluorescence microscopy to selectively examine the spatial distribution of negatively charged NVs upon local optical excitation of type-1b diamond. We find that carrier diffusion and subsequent trapping of photo-generated electrons and holes results in the emergence of singular patterns of trapped charge. In particular, we demonstrate that intense, fixed-point illumination at select wavelengths affects the NV⁻ concentration non-locally to produce regions of the crystal where the content of trapped charge is either enhanced or depleted relative to the background. Remarkably, the resulting NV⁻ distribution is mostly insensitive to the initial conditions, and arises from a subtle interplay between carrier photo-generation and trapping rates, electron and hole self-diffusion constants, and the type and concentration of atomic defects. We theoretically model the underlying dynamics and manage to qualitatively reproduce most observations.

## Results

*Optical control of NV charge*

The approach we follow in our experiments is illustrated in Fig. 1 (see also Supplementary Fig. 1). We use green (532 nm) and red (632 nm) laser light and a homemade fluorescence microscope to locally excite and subsequently image a [111], type-1b diamond crystal (40 ppm nitrogen content). Red excitation photo-ionizes the NV⁻ ensemble to produce neutral NVs (denoted below as NV⁰). It is generally accepted that this process — investigated at the single NV level[13,14] — takes place via a two-step absorption mechanism comprising the consecutive excitation of the excess electron into an intermediate, excited state and then into the conduction band (Fig. 1b and Supplementary Fig. 2); tunneling from an NV⁻ in the excited state to an adjacent nitrogen is also possible. On the other hand, neutral NV's are transparent to light at 632 nm (the zero-phonon-line is 575 nm), and hence the back transformation can only take place, e.g., via tunneling from the NV⁰ ground state or two-photon absorption (see below). The latter processes, however, are comparatively slow and therefore, red-induced fluorescence must gradually decay as an increasing fraction of negatively charged NVs transitions to the neutral state. This is precisely the observation in Fig. 1c where we plot the measured fluorescence as a function of time upon red illumination of variable intensity. Evidence of the red-induced conversion of NV⁻ into NV⁰ is presented in Supplementary Fig. 3.

[1]Dept. of Physics, CUNY-City College of New York, New York, NY 10031, USA. [2]CUNY-Graduate Center, New York, NY 10016, USA. [3]Laser Physics Centre, Research School of Physics and Engineering, Australian National University, Canberra, Australian Capital Territory 0200, Australia. *e-mail: cmeriles@ccny.cuny.edu

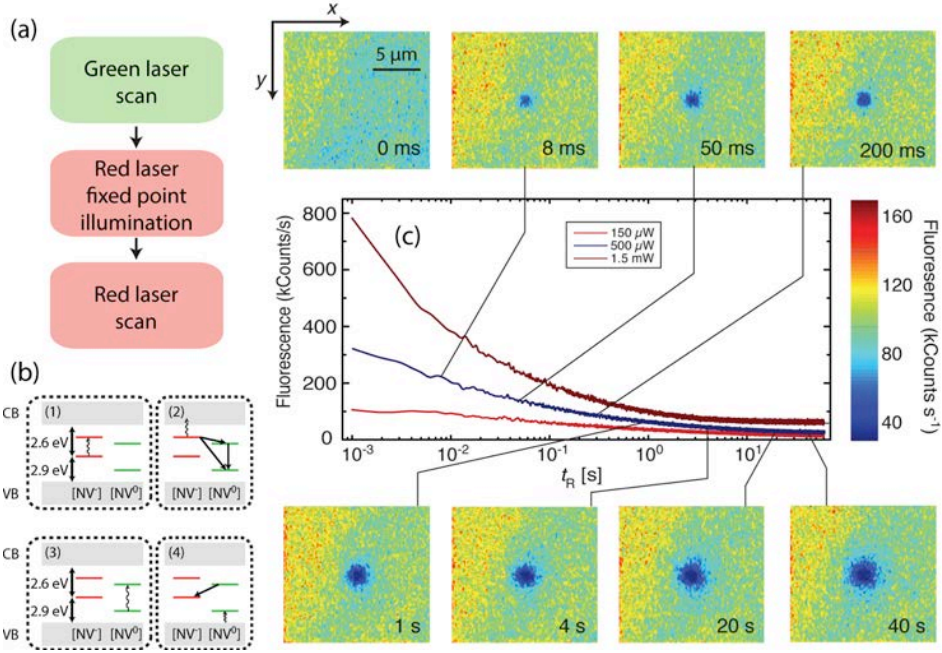

**Fig. 1 | Red-light-induced bleaching of the NV⁻ fluorescence.** (a) Experimental protocol. Upon a green laser sample scanning, we station the red beam at a fixed spot for a predefined, variable time, to finally image the area via a red beam scan. During the fixed-point illumination and scanning times the green beam is off. (b) NV ionization takes place via a two-step process in which the excess electron is ejected into the conduction band (CB) by the consecutive absorption of two photons of wavelength $\lambda \leq 632$ nm (steps 1 and 2). NV⁰ recharging takes place upon absorption of an electron from the valence band (VB) following excitation with $\lambda \leq 575$ nm (steps 3 and 4). (c) The central figure displays the sample fluorescence as a function of the red illumination time $t_R$ for three different excitation powers and prior to the readout scan. The top/down inserts correspond to fluorescence images obtained upon scanning the red beam over an area $xy$ on the focal plane around the fixed point. The time lapse $t_R$ prior to sample scanning is indicated in the lower right corner of each image. In all images, the green laser power is 750 µW; note that the red laser power is 500 µW during $t_R$ and 100 µW during the readout scanning, thus leading to different fluorescence intensities. Each image has 100×100 pixels and the integration time per point is 1 ms both throughout the scanned images and the fluorescence time traces in the central plot.

At low laser powers, NV ionization takes place on a slow time scale[14], hence allowing one to scan-image a vicinity of the excited area without considerably altering the local NV⁻ concentration. Images of the system fluorescence upon fixed-point red illumination of variable duration are presented as inserts in Fig. 1c. We witness the formation of a ~3-µm-wide dark area (slightly larger than the size of the red beam, see Supplementary Fig. 1) consistent with the progressive neutralization of negatively charged NVs within the excited volume. Note that unlike red light, green excitation dynamically interconverts the NV charge state from negative to neutral and vice versa (Fig. 1b and Supplementary Fig. 2). In the present case we exploit this feature to bring the NV⁻ content to a reproducible initial state, which we attain via a short, moderately intense green laser 'reset' scan preceding red illumination (Fig. 1a).

A subject of interest in the investigation of charge dynamics in diamond — largely overlooked so far — concerns the redistribution of photo-generated carriers during and after optical excitation. The experiment in Fig. 2 introduces a versatile platform to tackle this problem: Following green light scanning, we repeatedly bleach the area of interest via several consecutive red-laser sweeps. Images before and after red illumination (Figs. 2b and 2c, respectively) show a considerable reduction of the NV⁻ fluorescence, as anticipated from Fig. 1. We then park the red laser beam approximately at the centre of the scanned area for a predefined time $t_R$, to subsequently take an image via a 'readout' red laser scan. Figs. 2d and 2e show the results, both in the shape of a bright two-dimensional (2D) torus concentric with the laser excitation. The outer torus radius increases logarithmically with $t_R$, while the inner radius remains approximately constant (Fig. 2h).

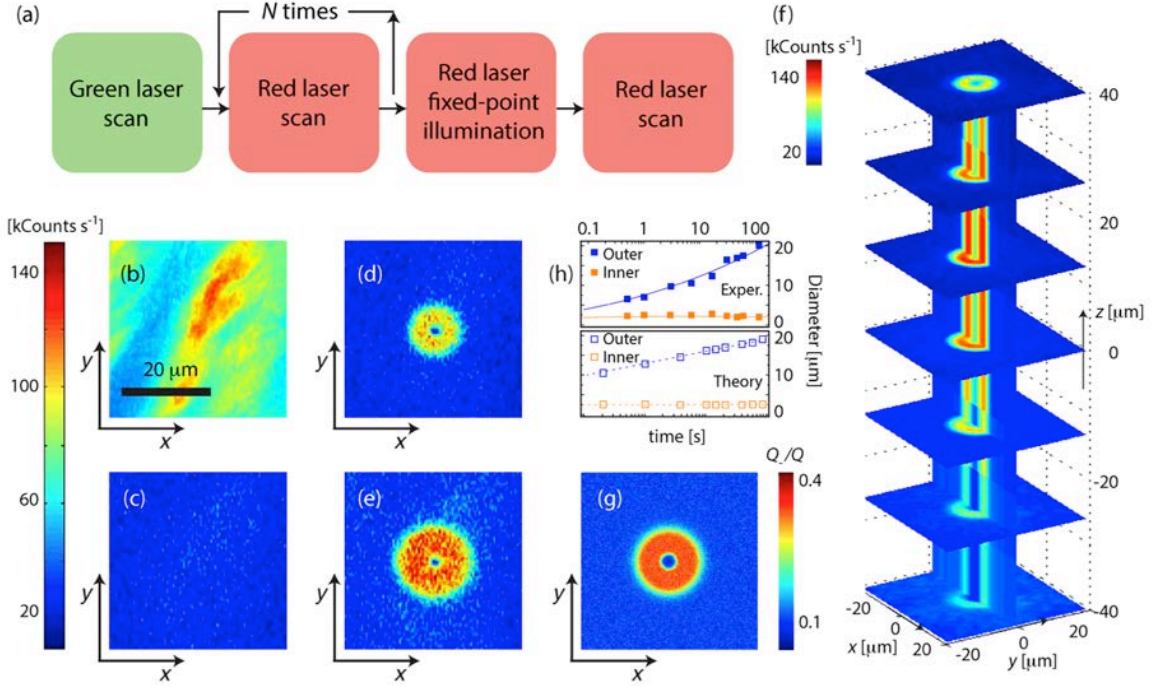

**Fig. 2 | NV⁻ formation during fixed-point, red-light excitation.** (a) Experimental protocol. A scanning green beam brings a predefined sample area to a well-defined reference configuration. Upon partially bleaching the sample through repeated red beam scanning, the red laser is focused on an arbitrary (but fixed) point for a time $t_R$. Finally, a fluorescence image is acquired via a readout scan. (b) Reference fluorescence image as obtained immediately after the green laser reset scan; the green beam power is 0.4 mW. (c) Fluorescence image after $N$=5 red laser bleach scans and in the absence of fixed-point illumination. (d,e) Fluorescence image after prolonged red excitation at the centre point of the scanned area. The red laser power is 2.0 mW during bleaching and fixed-point excitation, and 0.1 mW during scanning; the fixed-point excitation times are 15 s in (d) and 60 s in (e). (f) Three-dimensional reconstruction of the NV⁻ pattern resulting from 25 s of red, fixed-point excitation. All other conditions as in (d,e). (g) Numerical calculation of the map in (e) as determined from Eqs. (1-4) for 60 s red illumination (see (e) for comparison). (h) Time evolution of the outer and inner diameters of the observed (upper half graph) and calculated (lower half graph) bright 'torus'. Filled (empty) squares are experimental (calculated) data points; solid (dashed) lines serve only as a guide to the eye. In all scanned images the size is 100 x 100 pixels and the integration time per pixel is 1 ms. The scale bar in (b) applies to all images. See Supplementary Table 1 for a summary on the simulation parameters.

We interpret the formation of the bright torus-shaped pattern as the result of a light-assisted process in which neutral NVs in the near vicinity of the beam switch to negative upon capture of electrons diffusing away from the illuminated area at the torus centre[15]. The size of the bright area — much larger than the dark inner circle in Figs. 2d and 2e — suggests that defects other than the NV⁻ serve as the main source of electrons. The natural candidate is substitutional nitrogen, a deep donor and the dominant impurity in synthetic type-1b diamond, typically outnumbering NV centres by a factor of ten or more. Nitrogen is non-fluorescing and hence invisible in the scanned images of Fig. 2. Despite some initial controversy, recent experiments indicate that red excitation photo-ionizes substitutional nitrogen[16-19]. Therefore, multiple electrons can be generated in the illuminated area without substantially affecting the local number of negatively charged NVs, in agreement with our observations. Note that vacancies cannot be ionized by red (or green) excitation[20] and thus need not be considered in our experiments (see Methods and Supplementary Note 1)

*Modeling charge generation, diffusion and trapping*
To more quantitatively interpret our findings, we model the system dynamics via a set of coupled master equations for the local concentration of neutral and negatively charged defects in the presence of localized illumination and free diffusing photo-excited carriers (Supplementary Note 1). Denoting as $Q_-(\mathbf{r},t)$ the concentration of negatively charged NVs at position $\mathbf{r}$ and time $t$, we write

$$\frac{dQ_-}{dt} = -k_-Q_- + k_0Q_0 + \kappa_n n Q_0 - \kappa_p p Q_-, \quad (1)$$

where $k_- \equiv k_-(I(\mathbf{r}),\lambda)$ denotes the NV⁻ photo-ionization rate (in turn, a function of the laser intensity $I(\mathbf{r})$ and wavelength $\lambda$), $\kappa_n$ is the NV⁰ electron trapping rate, $Q = Q_- + Q_0$ is the total NV concentration, including the negative and neutral charge states, respectively, and $n \equiv n(\mathbf{r},t)$ is the concentration of conduction electrons. For reasons that will become apparent shortly, Eq. (1) also includes contributions from holes with concentration $p \equiv p(\mathbf{r}, t)$, generated by photo-excited, neutral NVs transforming to NV⁻ with rate $k_0$; correspondingly the last term takes into account contributions from holes trapping at NV⁻ sites with rate $\kappa_p$. The concentration of neutral nitrogen $P_0(\mathbf{r}, t)$ is governed by the analogous equation

$$\frac{dP_0}{dt} = -k_N P_0 + \gamma_n n P_+ - \gamma_p p P_0, \quad (2)$$

where $k_N = k_N(I(\mathbf{r}), \lambda)$ is the ionization rate of neutral nitrogen, $\gamma_n$ and $\gamma_p$ are the electron and hole trapping rates of neutral and positively-charged nitrogen, respectively, and $P = P_0 + P_+$ is the total nitrogen concentration. Finally, we describe the dynamics of photo-excited electrons and holes via the set of modified diffusion equations

$$\frac{dn}{dt} = D_n \left( \frac{\partial^2 n}{\partial r^2} + \frac{1}{r}\frac{\partial n}{\partial r} \right) - \kappa_n n Q_0 \\ - \gamma_n n P_+ + k_- Q_- + k_N P_0, \quad (3)$$

and

$$\frac{dp}{dt} = D_p \left( \frac{\partial^2 p}{\partial r^2} + \frac{1}{r}\frac{\partial p}{\partial r} \right) - \kappa_p p Q_- \\ - \gamma_p p P_0 + k_0 Q_0, \quad (4)$$

where $D_n \approx 5.5$ µm² ns⁻¹ and $D_p \approx 4.3$ µm² ns⁻¹ are the electron and hole diffusion coefficients in diamond[21,22], respectively. Throughout our calculations we assume cylindrical symmetry, a convenient simplification here justified by the poor localization of the laser beam along the axis of illumination (see Fig. 2f).

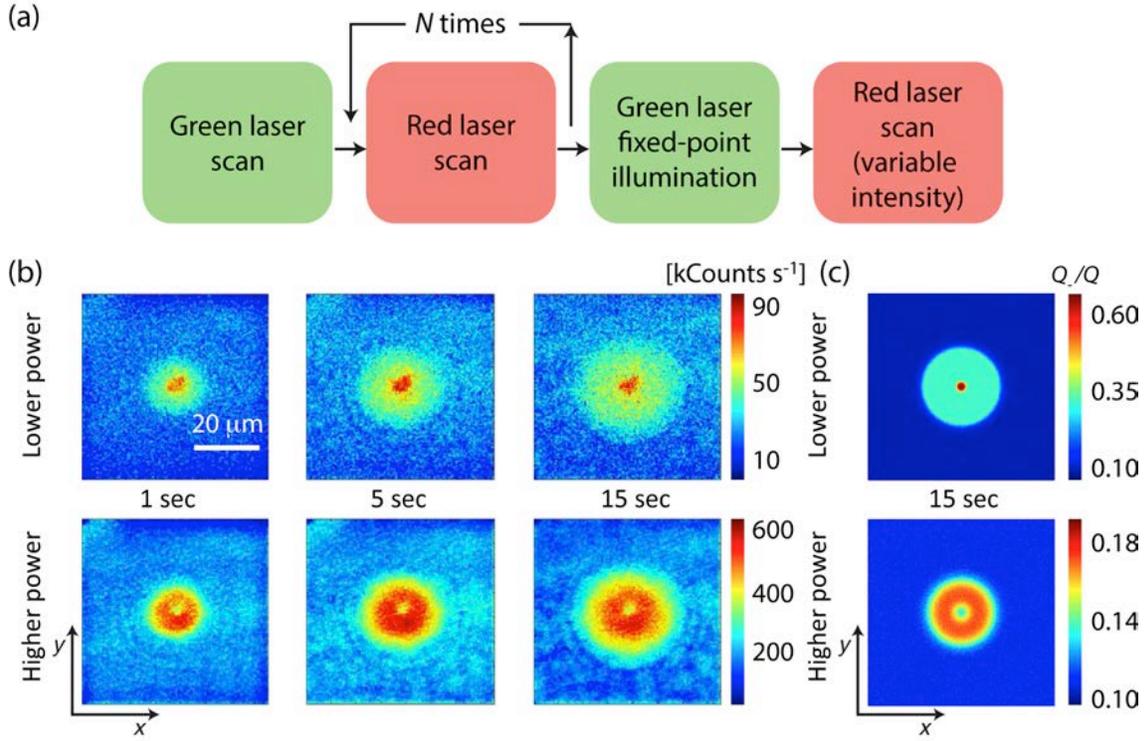

**Fig. 3 | Mapping NV⁻ patterns from fixed-point green-light excitation.** (a) Upon initializing the area via a strong green laser scan, we repeatedly scan the red beam to bleach the system fluorescence. We then point the green laser beam to a fixed point and illuminate for a fixed time interval $t_G$, to finally take a fluorescence image by scanning the red beam. (b) Fluorescence images upon a readout scan in an area around the point of green illumination (~1 µm diameter spot approximately at the centre). From left to right, the time $t_G$ of green laser illumination is 1 s, 5 s, and 15 s, respectively. Throughout the upper (lower) row of images, the red laser power during the readout scan is 100 µW (1.5 mW); in all cases the green laser power is 1.8 mW. All images share the same scale bar in the upper left and have a size of 150x150 pixels recorded with an integration time per pixel of 1 ms. (c) Calculated NV⁻ distribution taking into account the effect of the red laser beam during scanning (100 µW for the top image and 1.5 mW for the lower image). The scale bar and axes labels in (b) also apply to all images in (b) and (c). All parameters as listed in Supplementary Table 1.

A 2D plot of the calculated NV⁻ distribution arising from fixed-point, red illumination is presented in Fig. 2g. In our calculations we use the known electron and hole capture rates at nitrogen sites[21-24] ($\gamma_n \sim 200$ kHz µm³, and $\gamma_p \sim 0.6$ kHz µm³) and model[14] the NV⁻ photo-ionization rate at 632 nm as $k_- = k_-^{max}(f(r))^2$, where $k_-^{max}$ is a fitting parameter, $f(r)$ is the normalized (Gaussian) beam shape expressed as a function of the radial distance $r$, and we take into account the quadratic dependence with the illumination intensity[14]. Assuming[25] that a small fraction of the nitrogen atoms forms NVs ($Q = 10^{-2}P$, see Methods below), best agreement with our experimental observations is attained for the NV⁰ electron trapping rate $\kappa_n \sim 100$ kHz µm³, and $k_-^{max} \sim 4$ kHz; the ionization of nitrogen at 632 nm is modeled as $k_N = k_N^{max} f(r)$, with $k_N^{max} \sim 3 \times 10^{-2}$ Hz and we assume a one-photon process (see Figs. 2e and 2g). Our formalism captures the evolution of the bright toroidal shape with the duration of the fixed-point illumination interval (lower half of Fig. 2h) but we warn the agreement must be considered qualitative due to various simplifications in our model (see below, and Supplementary Notes 1 and 2).

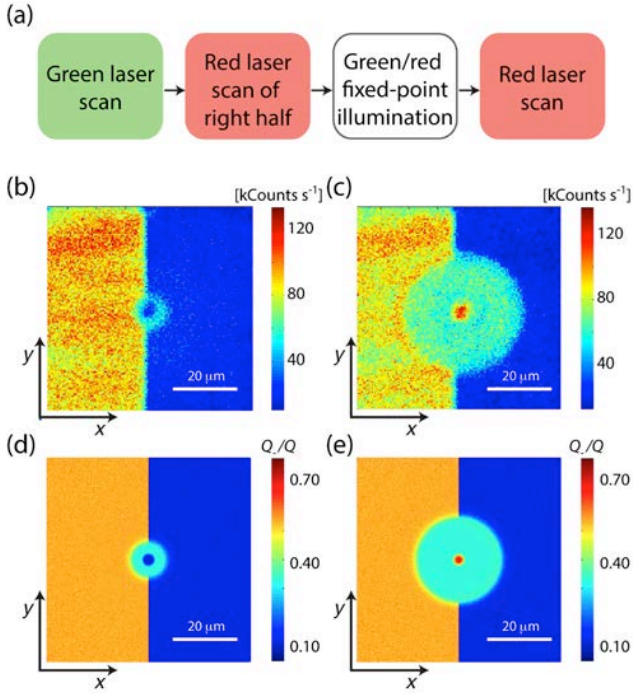

**Fig. 4 | Charge pattern formation in the presence of heterogeneous NV⁻ content.** (a) After initializing with a (green) reset scan, we ionize NV⁻ on half the sample area via a strong (red) bleach scan; we then focus either the red or green laser to a point approximately at the centre for a fixed time interval, and finally image the result via (a low intensity red) readout scan. (b) Fluorescence image upon application of the protocol in (a) using red, fixed-point excitation during $t_R$=10 s. The red beam power is 1.5 mW during the bleach scan, 1.5 mW during $t_R$, and 100 μW during the readout scan. (c) Same as in (b) but for green fixed-point illumination at 3 mW during $t_G$=10 s. In (b) and (c) the image size is 150x150 pixels and the integration time per pixel is 1 ms. (d,e) Calculated NV⁻ map for the conditions in (b,c), respectively. Simulation parameters as listed in Supplementary Table 1.

The NV⁻ pattern radically changes if green rather than red is used during the fixed-point illumination time. This is shown in Fig 3b (upper row) where the 2D NV⁻ distribution observed above now takes the form of a bright central spot surrounded by a weaker, time-dependent halo. Key to the formation of these patterns is the back-conversion of neutral NVs to the negatively-charged state upon optical absorption (Fig. 1b and Supplementary Fig. 2), a process that must be accompanied by a local enhancement of the NV⁻ fluorescence in the illuminated area as experimentally observed. Green excitation, on the other hand, is sufficiently energetic to ionize the NV⁻, implying that negative and neutral NVs are dynamically interconverted at a rate that sensitively depends on the illumination intensity[13,14]. The laser-induced cycle of NV charge and subsequent ionization is accompanied by the corresponding injection of holes and electrons into the valence and conduction bands, respectively. Consequently, the halo in the observed NV⁻ patterns emerges as some of the photo-generated carriers escape the area directly exposed to the laser. Note that diffusing holes and electrons undergo distinct dynamics: While the former are selectively captured by negatively charged NVs and neutral nitrogen, the latter can be trapped by positively charged nitrogen or by neutral NVs.

To quantitatively interpret our observations we adapt the set of Eqs. (1-4) to green excitation (Supplementary Note 2). The calculated result for fixed-point illumination of duration $t_G$ is presented in the upper graph of Fig. 3c, assuming that the photo-ionization rates of neutral and negatively charged NVs are identical. Best agreement with the experiment is attained using the nitrogen photo-ionization rate $k_N$~30 kHz and the NV⁻ hole capture rate $\kappa_p$~2$\kappa_n$~200 kHz μm³. Note that the condition $\kappa_n < \kappa_p$ can be anticipated since the NV⁰ is a neutral electron trap whereas the NV⁻ electrostatically attracts holes. On the other hand, we caution that the many degrees of freedom governing the system dynamics introduces considerable uncertainty in the numerical values extracted from our model, which, correspondingly, must be interpreted as a rough estimate (see Supplementary Note 2, and Supplementary Figs. 4 and 5).

The numerical results of Fig. 3c explicitly take into account the charge dynamics during the final readout scan required to form an image. This latter step is inconsequential when the illumination intensity is sufficiently weak because the scanning-induced NV⁻ ionization is negligible in this limit, meaning that the observed fluorescence count is directly proportional to the local NV⁻ concentration. The situation changes in the opposite regime of intense beam scanning, because the red-induced bleaching of the local NV⁻ fluorescence also depends on the charge state of nitrogen defects in the area: Faster NV⁻ bleaching rates are to be expected in parts of the crystal where most nitrogen impurities have been previously ionized because excess electrons can be removed more quickly from the group of illuminated NVs. This scenario is experimentally and numerically confirmed in the lower 2D graphs of Figs. 3b and 3c where the more intense readout scan makes the fluorescence patterns take an inverted shape, with a local minimum at the centre. We will show later that the notion of an environment-sensitive NV⁻ bleaching rate can be generalized to obtain a map of the positively charged nitrogen concentration.

A remarkable feature observed in the formation of these patterns is that the resulting charge distribution is mostly independent of the initial conditions. This behaviour is vividly demonstrated in Fig. 4 where we bleach only half of the scanned area and excite a point at the interface. We find that both for red (Fig. 4b) and green (Fig. 4c) fixed-point illumination, the emerging pattern is almost identical on both halves of the scanned area, despite the very dissimilar initial concentrations of NV⁻. In particular, we note that the left half of the red-induced 'torus' — brighter than the background in Figs. 2d and 2e — now takes the form of a dimmer half-ring. Something similar can be said about the green-induced halo of Fig. 4c, less bright than the background in the unbleached half of the crystal.

Both patterns can be reproduced reasonably well by solving the set of coupled differential equations presented above (Figs. 4d and 4e), although some subtle discrepancies hint at underlying dynamics somewhat overlooked by our model. Best agreement is attained in the case of green fixed-point excitation where the numerical simulation correctly produces a bright, ~3-μm-diameter disk surrounded by a flat halo of intermediate NV⁻ concentration on both halves of the imaged area (Fig. 4e). Photo-excited NV centres and nitrogen impurities play different roles in generating the observed pattern: Diffusing carriers produced by the former tend to lower the overall halo brightness because holes are preferentially attracted to NV⁻ centres whereas electrons tend to get trapped by positive nitrogen ions. It is the extra influx of electrons from photo-ionized nitrogen that somewhat increases the NV⁻ concentration in the halo to a level above that observed in the pre-bleached half.

The pattern in Fig. 4b is, however, more puzzling because solely electrons are (presumably) being generated by red, fixed-point excitation, which can only increase the NV⁻ concentration of the area surrounding the beam. This intuitive prediction contrasts with the experiment, and suggests the existence of concealed mechanisms of NV⁻ depletion, likely via the absorption of holes originating from a red-induced reconversion of NV⁰ into NV⁻ in the illuminated area. For example, good agreement with our observations can be attained by assuming only a low probability of reconversion of NV⁰ into NV⁻ (i.e., $(k_0 \cong 0.05 k_-)|_{\lambda=632\,nm}$, compare Figs. 4b and 4d). Note that even a small $k_0$ value can have a substantial effect given the relatively large hole capture rate of the NV⁻ compared to N⁰, at least in part, a consequence of the electrostatic attraction present only in the case of the NV⁻ centre (from Figs. 4c, 4e we find $\kappa_p \sim 330 \gamma_p \sim 200$ kHz μm³). We warn that the generation of holes could also originate from other defects not contemplated in our model. Besides some possible nitrogen complexes — likely coexisting with the more abundant substitutional nitrogen defect — one candidate is the silicon-vacancy (SiV) centre, present in this sample at a concentration of ~6 ppb (see Methods section and Supplementary Fig. 1). The SiV is known to exist in the negative and neutral charge states, with zero-phonon lines at 737 nm (1.69 eV) and 939 nm (1.32 eV), respectively[26,27]. Another possibility is the nitrogen-vacancy-hydrogen complex, a common defect in CVD diamond known to exist in the neutral and negatively-charged states[25]. Both for the SiV and the NVH centers, red (and green) illumination can, in principle, dynamically interconvert one charge state into the other with the corresponding additional production of electrons and holes. However, since the ionization rates of these defects are unknown, their impact on the observed dynamics is difficult to ascertain at this point.

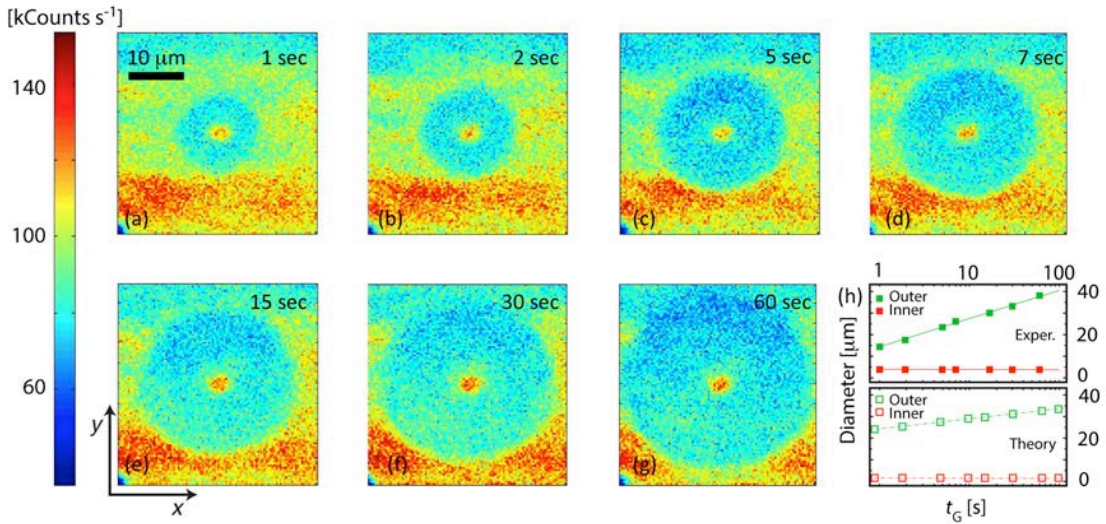

**Fig. 5 | Time evolution of the photo-induced NV⁻ pattern.** Following a green scan (reset), we use a weak red beam to image the pattern resulting from green, fixed-point illumination of variable duration $t_G$. Unlike the protocol in Fig. 3, no prior bleaching is carried out. The green, fixed-point illumination times are (a) 1 sec, (b) 2 sec, (c) 5 sec, (d) 7 sec, (e) 10 sec, (f) 15 sec, (g) 30 sec, and (h) 60 sec. The green and red laser powers are 1.8 mW and 100 μW, respectively. All images are recorded by scanning the red laser beam one time. All other conditions as in Fig. 4. (h) Observed (upper graph) and calculated (lower graph) dependence of the inner and outer diameters of the NV⁻ pattern as a function of $t_G$. Filled (empty) squares are experimental (calculated) data points; solid and dashed lines serve as a guide to the eye. The scale bar in (a) and axes labels in (e) apply to all scanned images. All parameters as listed in Supplementary Table 1.

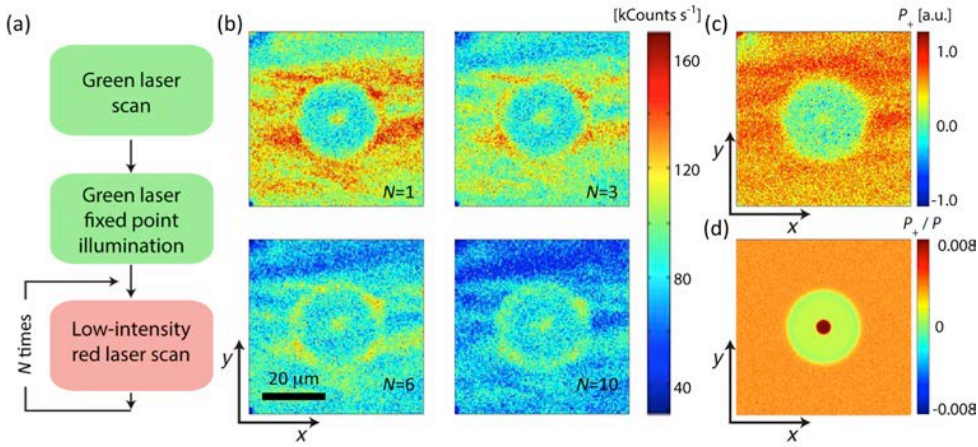

**Fig. 6 | Determination of the relative content of positively charged nitrogen.** (a) Experimental protocol. After green, fixed-point illumination, the sample is image-scanned $N$ times. (b) Resulting fluorescence images upon $N=1$ (top left), $N=3$ (top right), $N=6$ (bottom left), and $N=10$ (bottom right) readout scans. Besides an overall reduction of the sample fluorescence, successive red scanning alters the image contrast due to site-dependent bleaching rates. (c) Map of positively charged nitrogen $P_+$ as determined from the normalized bleaching rate at each point (see text). Most nitrogen impurities in the halo region are neutral. (d) Calculated $P_+$ two-dimensional map using the parameters listed in Supplementary Table 1. Images in (b) are 150x150 pixels and the integration time per pixel is 1 ms. The green and red laser powers are 1.83 mW and 110 µW, respectively. The green laser fixed-point illumination time is $t_G$=15 s. The scale bar in (b) applies to all images in (b) and (c).

The way these patterns evolve with time deserves special consideration. For example, Fig. 5 shows a sequence of images recorded upon green fixed-point excitation of variable duration $t_G$. Rather than a widening, Gaussian-like distribution, the halo region expands radially with a well-defined front separating different NV⁻ concentrations. The existence of a boundary suggests that an equilibrium concentration of negatively (positively) charged NVs (nitrogen) is established on a time scale shorter than that governed by carrier self-diffusion. This behaviour is analogous to that observed in reaction kinetics and is a direct consequence of the reaction-diffusion structure of the master equations governing the system dynamics. Note that the cylindrical symmetry in the patterns of Fig. 4 — insensitive to the initial NV⁻ content — can be interpreted along these same lines, since one would anticipate the equilibrium NV⁻ fraction to be an exclusive function of the ratio between the number of free electrons and holes reaching a given point, independent of the initial charge distribution between NVs and nitrogen centers. Phenomenologically (upper half in Fig. 5h), we find that the outer halo diameter $d_o$ grows with time as $d_o \sim \delta(I)\log(t/\tau(I))$, where $\delta(I)$ and $\tau(I)$ are functions of the laser intensity (see Supplementary Fig. 6). A similar response is predicted numerically for the present illumination conditions (lower half of Fig. 5h). We note, however, that the red-induced pattern follows a slightly different, non-logarithmic time dependence (Fig. 2h). The origin of this trend — growing with the applied laser intensity, see Supplementary Fig. 6 — is presently unknown, though a detailed analysis seems to rule out contributions from electrostatic fields, not taken into account by our model (see Supplementary Notes 1 and 2, and Supplementary Fig. 7).

In the typical regime where $Q \ll P$, the rate at which the NV⁻ ensemble bleaches upon red illumination is influenced by the average charge state of the surrounding nitrogen. For example, in the limit where all nitrogen impurities are positively charged, the bleaching rate of the illuminated NV⁻ ensemble is effectively faster because photo-generated electrons can be easily captured locally by immediately adjacent N⁺. The opposite is true when all nitrogen impurities are neutral because only self-diffusion outside the illuminated area can prevent electrons from being recaptured by neighboring NVs. Certainly, this picture rests on the assumption of a relatively low nitrogen ionization rate, a condition met here since $(k_N^{\max} \ll k_-^{\max})|_{\lambda=632\,nm}$ (Fig. 2). In the regime where the red scan intensity is sufficiently low, this connection can be exploited to determine the local (relative) fraction of positively charged nitrogen via the ansatz $P_+(\mathbf{r}) \propto (Q_-(\mathbf{r}))^{-1}|dQ_-/dt|\big|_\mathbf{r}$, where the last factor represents the effective NV⁻ bleaching rate at position $\mathbf{r}$ in the scanned area. It should be noted that NVs have served as a probe of optically-induced space charge distributions in diamond before, albeit using a different sensing mechanism (dc Stark shifts) and in high-purity material[28].

We implement the notion of NV-enabled $P_+$ sensing in Fig. 6 where we measure the bleaching rate at each point by comparing successive red-scan images of an area initially exposed to green fixed-point illumination. As expected, we find that $P_+(\mathbf{r})$ has a local maximum at the point of laser excitation — centre point in Fig. 6c — to subsequently decay in the surrounding vicinity, in agreement with the notion of positive nitrogen in the periphery preferentially capturing electrons photo-generated at the centre. Also worth highlighting is the background region (virtually unaltered by the green beam), where there is a correspondence between regions of high $P_+$ and $Q_-$ content, consistent with the idea of negatively charged NVs forming at the expense of nitrogen ionization. Overall, the observed pattern is in qualitative agreement with the calculated $P_+$ map (Fig. 6d), though some aspects of the dynamics at play are in need of further investigation. This is, for example, the case of the slowly bleaching ring around the 'halo' of depleted NV⁻ content, apparent in the lower images of Fig. 6b. Preferential electron trapping in this section of the crystal may be the manifestation of a pseudo-electrostatic, positive potential near the halo boundary. Further work, however, will be necessary to properly take into account this or similar inter-carrier effects, here ignored given the complexity of the accompanying set of master equations (Supplementary Note 1).

**Discussion**

In extending the experiments herein, the ability to control the spin of photo-generated carriers is particularly intriguing. Optical and microwave pulses can be easily articulated to project the negatively charged NV — a spin-1 system — into an arbitrary spin state. Therefore, if properly initialized (e.g., into the $m_S = 1$ level of the ground state), the NV⁻ could arguably be exploited as a resource for electron spin injection, provided ionization takes place on a time scale shorter than that inherent to the NV intersystem crossing (of order 1 μs). Whether the spin polarization of photo-generated carriers survives ionization, transport, and subsequent trapping (e.g., in a neighboring, neutral NV) as indirectly suggested by recent observations[13] is a question of both fundamental, and practical interest, which, surprisingly, has not yet been explored. For example, spin-polarized electrons propagating from a source to a target NV could serve as a quantum bus to mediate the interaction between remote, in-chip spin qubits[12]. On the other hand, the use of diamond as a material platform for quantum spintronics is particularly attractive because its large electronic bandgap, inversion symmetry, small spin-orbit interaction, and low nuclear spin density promise long carrier spin relaxation times, even at room temperature[29].

Systematic observations over a period of a week suggest that, once created, these patterns persist indefinitely in the absence of optical illumination. The latter is consistent with the relatively large gap separating the ground state of NV centers and nitrogen impurities from the conduction band minimum (2.6 eV and ~1.7 eV, respectively) and suggests that charge tunneling between defects is limited at the present concentrations. The ability to store, read, and rewrite classical information in the form of local patterns of trapped charge may be exploited to develop compact, three-dimensional memories, especially if super-resolution or multi-photon excitation techniques manage to bring the volume per bit below the light diffraction limit.

**Methods**

*Experimental setup and sample characterization*

All experiments are carried out using a custom-made, two-color microscope as sketched in Supplementary Fig. 1. Excitation in the red (632 nm) or green (532 nm) is provided by a 13 mW helium-neon laser or a 2 W cw solid-state laser, respectively. The illumination timing is set independently with the aid of acousto-optic modulators (AOM); a servo-controlled, two-mirror galvo system is used for sample scanning. Sample fluorescence in the range 650 – 850 nm is detected after a dichroic mirror and notch filters by a solid-state avalanche photo detector (APD). A 0.42 numerical-aperture (NA) objective is used to focus either beam on the sample and collect the outgoing fluorescence. All experiments are carried out under ambient conditions.

To determine the laser beam shape $f(r)$ at the focal point we reconstruct the image formed by a beam reflecting from a silver mirror placed at the sample position. To this end we alter the green laser path so as to bypass the galvo (see Supplementary Fig. 1a); the beam shape is measured as the galvo scans the focal plane (set to coincide with the mirror surface). The resulting beam shape — shown in Supplementary Fig. 1b — has a nearly-ideal Gaussian profile, with a full width at half maximum (FWHM) of about 1 μm.

The sample used throughout our experiments is a type-1b diamond from Diamond Delaware Knives (DDK). Infrared spectroscopy (Supplementary Fig. 1c) is consistent with the presence of substitutional nitrogen but a quantitative determination is complicated by the overlap of the characteristic N⁰ peak at 1132 cm⁻¹ with other broad adjacent lines. An estimate using the method in Ref. 30 yields $P_0 = 40 \pm 5$ ppm, with the exact value depending on the deconvolution protocol. The low infra-red absorption near 1282 cm⁻¹ suggests that A-centres — formed by two adjacent nitrogens — are, if at all present, at trace concentrations[31]. Optical spectroscopy confirms that the collected fluorescence originates almost exclusively from NV centres (Supplementary Fig. 1d). A distinctive peak superimposed on the red-shifted NV fluorescence emission at ~737 nm reveals the presence of negatively charged silicon-vacancy (SiV) centres; from the peak amplitude we estimate the SiV-NV ratio to be about 0.6 %. Note that the absence of a peak at 741 nm indicates that most vacancies are, if present, in the negatively-charged state (with zero phonon line at 394 nm) and hence are immune to red or green illumination[20,32]. Throughout our modeling we set the NV concentration to $Q = 0.4$ ppm. This value — derived from the observed ratio $Q \sim 10^{-2} P$ between nitrogen impurities and naturally occurring NV centers in CVD diamond[25] — also provides the best agreement between experiment and modeling (see Supplementary Note 2).

During the experimental protocols of Figs. 1 through 6 the typical duration of the reset and imaging scans is 30 s. To create an NV⁻-depleted area, the red laser (1.5 mW) is typically scanned multiple times so as to bring the observed fluorescence to a minimum. The time lapse between steps typically ranges from 1 to 30 s depending on the particular protocol. Note that the exact wait time between steps is not crucial because all NV⁻ patterns were observed to remain unchanged in the dark for a virtually arbitrary time (days).

*Numerical simulation*

Given the cylindrical symmetry of the observed patterns, we restrict our simulations to just the radial dimension. To further simplify the equations, we assume that free carrier self-diffusion prevails over the electric forces generated by the separation of charge (e.g., $\mu_n \nabla \cdot (n\boldsymbol{E}) \ll D_n \nabla^2 n$, where $\mu_n$ is the electron mobility), such that electric forces can be neglected.

In each simulation, the model equations are solved over a $100 \times 100$ μm² region for a period extending up to 100 s. These dimensions are chosen so as to enclose the observed patterns and to satisfy the requirements of the assumed boundary conditions (see Supplementary Note 1). To recreate the effect of strong red laser scanning in Fig. 3c computations are carried out in two stages: During the first one, the system evolves in the presence of green fixed-point illumination from a homogeneously low NV⁻ distribution. The resulting NV⁻ pattern is then subjected to a (virtual) red beam of uniform intensity throughout the simulated sample area.

The model equations are solved using the numerical differential equation solver (NDSolveValue) of *Mathematica*. The simulations are repeated as the model parameters are swept logarithmically in the region of their expected values. The simulations that qualitatively compare the best with our observations are selected for presentation. The parameters of the selected simulations are detailed in Supplementary Table 1 and a discussion on the model accuracy is presented in Supplementary Note 2.

**Acknowledgements**

We thank Michael Barson and Halley Aycock-Rizzo for their assistance with some of the experiments and sample preparation. H.J., J.H., S.D., D.P., A.L., and C.A.M. acknowledge support from the National Science Foundation through grant NSF-1314205. M.D. and N.M. acknowledge support from the Australian Research Council through grant DP120102232.

**Data Availability**

The authors declare that the data supporting the findings of this study are available within the article and its supplementary information files.

**Author contributions**

H.J., J.H., and S.D. conducted the experiments, with the assistance of R.A., D.P., and A.L.; M.D. and S.D. developed the model for charge transport and conducted the numerical simulations. C.A.M., H.J., M.D., and R.A. conceived the experiments; C.A.M., M.D., and N.B.M. supervised the work. All authors analyzed the data and contributed to the writing of the manuscript.